\documentclass[aoas,MSNbibl,nameyear,dvips]{arximspdf}
\usepackage{graphicx}

\doi{10.1214/12-AOAS576} %kopijuoti is PTS
\volume{6}
\issue{4}
\pubyear{2012}
\firstpage{1478}
\lastpage{1498}

\makeatletter

\newcommand{\eqref}[1]{(\ref{#1})}
\newcommand{\btheta}{\bolds{\theta}}
\newcommand{\bmu}{\bolds{\mu}}
\newcommand{\bSigma}{\bolds{\Sigma}}
\newcommand{\brho}{\bolds{\rho}}

\newproclaim{definition}{Definition}[section]

\makeatother

\begin{document}
\begin{frontmatter}

\title{Spatial analysis of wave direction data using wrapped Gaussian processes}
\runtitle{Spatial directional data}

\begin{aug}
\author[A]{\fnms{Giovanna} \snm{Jona-Lasinio}\corref{}\ead[label=e1]{giovanna.jonalasinio@uniroma1.it}},
\author[B]{\fnms{Alan}~\snm{Gelfand}\ead[label=e2]{alan@stat.duke.edu}\thanksref{t1}}
\and\break
\author[C]{\fnms{Mattia}~\snm{Jona-Lasinio}\ead[label=e3]{mattia.jonalasinio@itp.uni-hannover.de}}
\thankstext{t1}{Supported in part by NSF Grants DMS-09-14906 and CDI 0940671.}

\runauthor{G. Jona-Lasinio, A. Gelfand and M. Jona-Lasinio}
\affiliation{University of Rome Sapienza, Duke University and Leibniz
University}
\address[A]{G. Jona-Lasinio\\
DSS University of Rome Sapienza\\
P. le Aldo Moro 5\\
00185 Rome\\
Italy\\
\printead{e1}}
\address[B]{A. Gelfand\\
Duke University\\
223-A Old Chemistry Building\\
Box 90251\\
Durham, North Carolina 27708-0251\\
USA\\
\printead{e2}}
\address[C]{M. Jona-Lasinio\\
ITP---Leibniz Universitat Hannover\\
Appelstrasse 2\\
30167 Hannover\\
Germany\\
\printead{e3}}
\end{aug}

% HISTORY:
\received{\smonth{10} \syear{2011}}
\revised{\smonth{5} \syear{2012}}

% ABSTRACT
%
\begin{abstract}
Directional data arise in various contexts such as oceanography (wave
directions) and meteorology (wind directions), as well as with
measurements on a periodic scale (weekdays, hours, etc.). Our
contribution is to introduce a model-based approach to handle periodic
data in the case of measurements taken at spatial locations,
anticipating structured dependence between these measurements. We
formulate a wrapped Gaussian spatial process model for this setting,
induced from a customary \textit{linear} Gaussian process.

We build a hierarchical model to handle this situation and show that
the fitting of such a model is possible using standard Markov chain
Monte Carlo methods. Our approach enables spatial interpolation (and
can accommodate measurement error). We illustrate with a set of wave
direction data from the Adriatic coast of Italy, generated through a
complex computer model.
\end{abstract}

% KEYWORDS
%

\end{frontmatter}

%s1 #&#
\section{Introduction}\label{sec1}

Directional or angular data arise, for instance, in ocea\-nography (wave
directions), meteorology (wind directions) and biology (study of animal
movement). They also arise from periodic data, for example, event times
might be wrapped to a weekly period to give a circular view
(eliminating end effects) of the pattern of event times. Here, we
assume the data is recorded in degrees or angles on a circle. This is
not a limitation, as any circular scale (e.g., $[0,L)$ or $[-L/2,L/2)$)
can be transformed to $[0,2\pi)$ by a modulus transformation. Handling
such data creates difficulties due to the restriction of support to the
unit circle, $[0,2\pi)$, and to the sensitivity of descriptive and
inferential results to the starting point on the circle. Hence,
analysis of directional data is more challenging than for linear data.
There exists a substantial literature on circular data [see, e.g.,
\citet{mardia}, \citet{mardiajupp}, \citet{Jamma} or \citet{fisher}], but,
broadly, it is confined to descriptive statistics and limited inference
for simple univariate models.

The contribution of this paper is to introduce a fully model-based
approach, that is, Bayesian hierarchical modeling, to handle angular
data, enabling full inference regarding all model parameters and
prediction under the model. Our focus is on multivariate directional
observations arising as angular data measurements taken at spatial
locations, anticipating structured dependence between these
measurements. Thus, we formulate an attractive spatial process model
for directional data, the wrapped Gaussian process, induced from a
linear (customary) Gaussian process. We illuminate the dependence
structure. We show how to implement kriging of mean directions and
concentrations in this setting. We work within a hierarchical Bayesian
framework and show that introduction of suitable latent variables
facilitates Markov chain Monte Carlo model fitting. We offer an
adaptive truncation strategy for simulation of these latent variables.

Directional data has a long history. Early contributors to the
theoretical development include Watson and Stephens [\citet{watson},
\citeauthor{stephens63} (\citeyear{stephens63,stephens70})]. \citet{kent78} studied complex
circular distributions. The books of \citet{mardia} and \citet
{mardiajupp} present approaches, distribution theory and inference for
such data. In \citet{fisher} we find comprehensive discussion, with
particular attention to nonparametric methods. Computational procedures
such as MCMC methods and the EM algorithm have enabled analysis for
directional data to become less descriptive and more inferential.
Examples include linear models
[\citet{harrison},
\citet{fisher},
\citet{fisher-lee92},
\citet{kato08}], linear models in a Bayesian
context
[\citet{guttorp},
\citet{damien-walker}] and models for circular time series
[\citet{breckling},
\citet{coles98},
\citet{mardiajupp},
\citet{ravindran},
\citet{hughes},
\citet{fisher-lee},
\citet{holtzmanetal06}].
Recently, Kato [\citet{kato10}], building upon earlier
work [\citet{kato08}], has proposed a discrete time Markov process for
circular data using the M\"{o}bius circle transformation, connecting it
with an early Markov process model of Fisher and Lee (\citeyear{fisher-lee}).
We offer a process model for locations in $d$-dimensional space but focus
on the 2-dimensional case.

There is little in the way of formal multivariate theory for circular
data, particularly in the fully Bayesian setting. In this regard,
perhaps the work of \citet{coles98} is the closest to ours. He also
employs wrapped distributions, noting that, in the Gaussian case, they
can be readily given a multivariate extension. Coles mostly works with
independent replicates of multivariate circular data in low dimension
with an unknown covariance matrix and develops some theory and examples
for the time series setting. However, he mentions possible extensions
to the spatial setting but offers no development, in particular, no
thoughts on regression or kriging (Sections~\ref{regWN} and~\ref{krigtheory} below). \citet
{casson-coles98} include spatial dependence in looking at the direction
of maximum wind speed. With little detail, they propose conditionally
independent directions modeled with a von Mises distribution,
introducing spatial structure in the modal direction and concentration
parameters. We introduce spatial structure directly on the angular
variables, with a single, potentially high-dimensional multivariate
observation but driven by a spatial process model, yielding a
high-dimensional covariance matrix with structured dependence as a
function of perhaps two or three parameters.

%coming, I found outgoing wave direction and incoming wave direction as
%definitions. Hopefully this sounds right. (2) Shall we mention here
%the fact that wind and wave directions have a conventionally fixed
%origin of the scale (0=North)? }

Our motivating example is drawn from marine data. Often, such data are
based on outputs from deterministic models, usually climatic forecasts
computed at several \textit{spatial} and \textit{temporal} resolutions.
Wave heights and outgoing wave directions, the latter being measured in
degrees relative to a fixed orientation, are the main outputs of marine
forecasts. Numerical models for weather and marine forecasts need
statistical post-processing; wave heights, like wind speed, being
linear variables, can be treated in several ways [\citet{kalnay},
\citet{wilks06},
\citet{jonaetal07}]. Wave directions, being angular variables,
cannot be
treated according to standard post-processing techniques [see
\citet{engel07},
\citet{bao10} and references therein]. In \citet{bao10} bias
correction and ensemble calibration forecasts of surface wind direction
are proposed. The authors use circular--circular regression as in \citet
{kato08} for bias correction and Bayesian model averaging with the von
Mises distribution for ensemble calibration. However, their approach
does not explicitly account for spatial structure.
In our setting, wave direction data is viewed differently from wind
direction data. The former is only available as an angle, while the
latter is customarily associated with wind speed, emerging as the
resultant of North--South and East--West wind speed components.

Eventually, we plan to do joint spatio-temporal modeling of wave height
and wave direction (linear and circular data), fusing numerical model
output with observed buoy data. As a first step, here, we take up
static spatial modeling for the WAve Model (WAM)
%in Europe is the European Center for Medium-Range Weather Forecasts
%(ECMWF), which runs at global medium range (3-5 up to 10 day
%forecasts, 55 km spatial resolution) and at high resolution short term
%(3 days, 25 km resolution Wave Amplitude Model, WAM) models in the
%Mediterranean Area.}
data (see Section~\ref{sectionrd}), deferring the joint modeling,
dynamics and data fusion for a future paper.
%The static spatial version has its own relevance as the need of
%downscaling wave direction data arise in coastal studies.
%Data are generated through a complex computer model (WAMl). These data
%are produced on a grid with $25\times25$ km cells and are related to
%wave height forecast in deep waters (more than 100m depth). In order
%to include directional information in shallow waters forecast models
%it is necessary to align these values with other linear data (wave
%heights, bathymetry, wind speed etc.) given at a higher spatial
%resolution ($10\times10$ Km grid).

The format of the paper is as follows. Section~\ref{univ} reviews wrapped
distributions for univariate circular data. Section~\ref{WGP} moves on to the
wrapped Gaussian process. Section~\ref{examples} takes up the wave direction
application and Section~\ref{sec5} offers a summary and future directions.

%s2 #&#
\section{Univariate wrapped distributions}\label{univ}

The von Mises (or circular normal) distribution [see \citet
{mardia},
\citet{mardiajupp}] is the most common specification for the
univariate case. It is extensively studied and inference techniques are
well-developed, but a multivariate extension is still a work in
progress. A few recent papers
[\citet{mardia-taylor07},
\citet{mardia-taylor08}] report applications of
bivariate and
trivariate von Mises distributions, with procedures that are complex
and computationally intensive. However, for usual spatial
settings, multivariate distributions of
dimension $50$, $100$ or more arise.

For the wrapping approach, let $Y$ be a random variable on
$\mathbb{R}$, henceforth
referred to as a \textit{linear} random or unwrapped variable, with probability
density function $g(y)$ and distribution function $G(y)$. The induced wrapped
variable ($X$) of period~$2\pi$ is given by
%
%e1 #&#
\begin{equation}
X=Y \operatorname{mod}2\pi. \label{eqwrapprocedure}
\end{equation}
Evidently, $0 \leq X < 2\pi$. The associated \textit{directional} probability
density function $f(x)$ is obtained by wrapping $g(y)$, via the transformation
$Y$=$ X+2K\pi$, around a circle of unit radius, with K being the
\textit
{winding number}. It takes
the form of a doubly infinite sum,
%
%e2 #&#
\begin{equation}
f(x)=\sum_{k=-\infty}^{\infty}{g(x+2k\pi)}, \qquad 0 \leq x<2
\pi. \label{eqEq145}
\end{equation}
%
% with distribution function,$F(x)=\sum_{k=-\infty}^{\infty
%}{G(x+2k\pi)-G(2k\pi)}, 0 \leq x<2\pi$.
From (\ref{eqEq145}), we see that the joint distribution of $(X,K)$ is
$g(x+2k\pi)$ with $x\in[0, 2\pi)$ and $k \in\mathbb{Z}\equiv\{0,
\pm
1, \pm2,\ldots\}$. That is, $Y \Leftrightarrow(X,K)$; $Y$ determines
$(X,K)$ and vice versa. Then,
marginalization over $k$ produces (\ref{eqEq145}).
%Expressed differently,
%$g(\cdot)$ is the distribution of $Y=X+2K\pi$, $Y$ determines $X$ and
%$K$ through \eqref{eqwrapprocedure}, and $X$ is a wrapped version of
%$Y$.
From this joint distribution the marginal distribution
of $K$ is $P(K=k) = \int_{0}^{2\pi} g(x+2k\pi) \,dx$. Additionally,
$K|X=x$ is
such that $P(K=k|X=x) = g(x+2k\pi)/\break\sum_{j= -\infty}^{\infty}
g(x+2j\pi
)$ while
the conditional distribution of $X|K=k$ is $g(x+2k\pi)/\int_{0}^{2\pi}
g(x+2k\pi) \,dx$. Hence, the wrapped distributions are easy to work with,
treating $K$ as a latent variable. In the context of simulation-based model
fitting, sampling the full conditional distribution for $K$ will be required.
The end of Section~\ref{wnd} provides an automatic truncation approximation to
facilitate such sampling.

Expectations under $f(x)$ are generally difficult to calculate;
%For instance, the mean $E(\cdot)$ of a function $h(x)$ of an angle in
%$ [0,2\pi)$ is
%E(h)=\sum_{k=-\infty}^{\infty}\int_{2k\pi}^{2(k+1)\pi}g(x)h(x-2k
a more convenient variable to work with is the associated complex
random variable on the unit circle in the complex plane,
$Z=e^{iX}$. In particular, for integer $p$, $E(e^{ipx}) = \psi_{Y}(p)$,
where $\psi$ is the characteristic function of $Y$ [\citet{Jamma}].

%s2.1 #&#
\subsection{The wrapped normal distribution}\label{wnd}

Under \eqref{eqEq145}, the wrapped normal distribution arises with
$g$ a normal density indexed by parameter $\btheta$ consisting of mean,
$\mu$, and variance $\sigma^{2}$. In fact, we can envision $\mu= \tilde{\mu} + K_{\mu}$, where
$\tilde{\mu}$ is the mean direction, a parameter on $[0,2\pi)$ and
$K_{\mu}$ is the associated winding number; as above, $\mu$ determines
$\tilde{\mu}$. We can reparametrize $\sigma^2$ to $c =e^{-\sigma
^{2}/2}<1$ where $c $ is referred to as the concentration parameter
[\citet{Jamma}, pages 27--28]. We write the wrapped normal distribution
of $X$ as $\operatorname{WN}(\mu,\sigma^2)$ with the probability density
function,
%
%e3 #&#
\begin{equation}
f(x)=\frac{1}{\sigma\sqrt{2\pi}}\sum_{k=-\infty}^{\infty
}{\operatorname{exp}
\biggl\{ \frac{-(x-\mu+2k\pi)^{2}}{2\sigma^{2}} \biggr\} },\qquad 0 \leq x<2\pi. \label{eqEq149}
\end{equation}

%
%Due to the form in \eqref{eqEq149}, we shall see that hierarchical
%model fitting is readily implemented with the WN distribution. In
%fact, it becomes an attractive alternative to the von Mises when we
%move to the multivariate setting.

From above, $E(Z^{p})=\int_{0 }^{2\pi}e^{ipx}f(x)\,dx=e^{ip\mu
-p^{2}\sigma^{2}/2}$.
Thus, we have
%
%e4 #&#
\begin{equation}
E(Z)=e^{-\sigma^{2}/2}(\cos\mu+i\sin\mu) \label{trigonom1}
\end{equation}
so that the resultant length of $Z$ is the foregoing concentration
parameter $c$. Furthermore, if $E\operatorname{cos}x = c \operatorname
{cos}\mu= c \operatorname
{cos}\tilde{\mu}$ and $E\operatorname{sin}x = c \operatorname
{sin}\mu= c \operatorname
{sin}\tilde{\mu}$, with $\tilde{\mu}$ the mean direction, as above,
then $\tilde{\mu} = \operatorname{arctan}^{*}(E\operatorname
{sin}X,E\operatorname
{cos}X)$.\setcounter{footnote}{1}\footnote{From \citeauthor{Jamma} [(\citeyear{Jamma}), page 13] $\operatorname
{arctan}^{*}(S,C)$ is formally defined as $\operatorname{arctan}(S/C)$
if $C>0,
S \geq0$; $\pi/2$ if $C=0, S > 0$; $\operatorname{arctan}(S/C)+ \pi
$ if $C<0$;
$\operatorname{arctan}(S/C)+2\pi$ if $C \geq0, S < 0$; undefined if
$C=0, S = 0$.}

To implement Markov chain Monte Carlo model fitting with wrapped normal
distributions, we introduce $K$ as a latent variable (Section~\ref{modfitind}).
Hence, we will have to sample $K$'s, one for each location, at each
iteration. It is difficult to sample over the support $ \{0, \pm1, \pm
2,\ldots\}$. However, it is well known that (\ref{eqEq149}) can be approximated with
only a few terms. For instance, \citet{mardiajupp} comment that, for
practical purposes, the density can be approximated by truncation to $k
\in\{-1,0,1\}$ when $\sigma^2 \geq2\pi$, while for $\sigma^2 <
2\pi$
using only $k=0$ gives a reasonable approximation. We can be more
precise. Suppose we
translate from $X$ to $X^{\prime}=(X + \pi)\operatorname{mod}2\pi- \pi$
to achieve
symmetric support,
$[-\pi,\pi)$, with corresponding translation of $\mu$ to $\mu^{\prime}$.
Then, suppressing the primes for convenience, with $\varphi$ denoting
the unit normal density
function,
\begin{eqnarray*}
\int_{-\pi}^{\pi}\sum_{k=-\infty}^{\infty}
\frac{1}{\sigma} \varphi\biggl(\frac{x+2k\pi- \mu}{\sigma}\biggr) \,dx&=& \sum
_{k=-\infty
}^{\infty} \int_{-\pi}^{\pi}
\frac{1}{\sigma} \varphi\biggl(\frac
{x+2k\pi-
\mu}{\sigma}\biggr)\,dx
\\
&=& \sum_{k=-\infty}^{\infty} \int
_{{((2k-1)\pi-\mu)}/{\sigma}}^{{((2k+1)\pi-\mu
)}/{\sigma}} \varphi(z) \,dz.
\end{eqnarray*}
Careful calculation
reveals that, if $k_{U} = 1+\lfloor\frac{3\sigma}{2\pi}\rfloor=
-k_{L}$ (where $\lfloor a\rfloor$
denotes the integer nearest to $a$ rounded toward 0), then
$(2k_{U}+1)\pi- \mu> 3\sigma$ and
$(2k_{L}-1)\pi- \mu< -3\sigma$. As a result,
%
%e5 #&#
\begin{eqnarray}\label{rule1}
\sum_{k=-\infty}^{\infty} \int_{{((2k-1)\pi-\mu)}/{\sigma}}^{({(2k+1)2\pi-\mu
})/{\sigma}}
\varphi(z) \,dz& > &\sum_{k=k_{L}}^{k_{U}} \int
_{{((2k-1)\pi-\mu)}/{\sigma}}^{{((2k+1)\pi-\mu
)}/{\sigma}} \varphi(z)\,dz
\nonumber
\\[-8pt]
\\[-8pt]
\nonumber
& > &\int
_{-3}^{3} \varphi(z)\,dz=0.997.
\end{eqnarray}

Expression \eqref{rule1} facilitates MCMC model fitting since it allows
us to determine the number of terms needed for good approximation as a
function of $\sigma$, for example,
% represents one of the main results of the present work. It allows us
%to determine the number of terms we need to well
%approximate a univariate normal wrapped distribution, providing at the
%same time an analytic upper bound for the truncation error. Another
%relevant advantage of this truncation is that it allows, in McMC
%algorithms, to choose $m$ adaptively at each step according to the
%current value of the generated $\sigma^2$, while keeping the
%approximation precision constant at each iteration.
%Using the above rule for $k_{U}$ and $k_{L}$,
if $\sigma< 2\pi/3$, then $k \in\{-1,0,1\}$; if $2\pi/3 \leq\sigma<
4\pi/3$, then $k \in\{-2,-1,0,1,2\}$.
%An analogous result can
%be obtained for the multivariate wrapped Gaussian distribution.
%However, we only
%need \eqref{rule1} since, for the model fitting under the wrapped
%Gaussian
%spatial process, we only simulate from univariate full conditional
%distributions
%for $K$.

%Inference for angular data requires care. For instance, if much of the
%data is near $0$ and $2\pi$, a sample average will be near $\pi$ and
%will not be meaningful. This emphasizes the need to work with a
%parameter
%requires a prior that is an angular distribution. Additionally,
So, $K$ can be large if and only if $\sigma^2$ can be large. Under a
simulation-based model fitting the pair will not be well-identified
unless we introduce an informative prior on $\sigma^2$.
Moreover, when the WN concentration $c$ is small ($\sigma^2$ large), it
becomes difficult to discriminate the WN from the uniform circular
distribution. With simulation experiments generating 1000 samples from
WN's using several combinations of sample sizes and variance values,
uniformity tests such as Rayleigh, Kuiper--Watson and Rao fail
to discriminate between the WN and the uniform distribution for $\sigma^{2}=3.252$
with small sample sizes ($n=30$), for $\sigma^2=4.02$ when
$n=100$ and $\sigma^2=7.11$ when $n=1000$. Hence, it will be worthwhile
to employ exploratory data analysis, for example, through the foregoing
tests (available,
e.g., in the CircStats library of R), with moments estimators
for $\tilde{\mu}$ and $\sigma^{2}$, in order to assess whether to model
using a WN. We recall these moments estimators: with angular data
$x_{1},\ldots
,x_{n}$ and the foregoing notation, let $\bar{C}=\frac{1}{n}\sum_{i=1}^{n}\cos x_{i}$ and $\bar{S}=\frac{
1}{n}\sum_{i=1}^{n}\sin x_{i}.$ Setting $\bar{C} = \hat{c}
\operatorname{cos}
\hat{\tilde{\mu}} $ and $\bar{S} = \hat{c} \operatorname{sin}
\hat{\tilde{\mu
}}$, we obtain moments estimators for $c$
and $\tilde{\mu}$ as
$e^{-\hat{\sigma}^{2}/2} = \hat{c} =\sqrt{\bar{C}^{2}+\bar{S}^{2}}$ and
$\hat{\tilde{\mu}} =\operatorname{arctan}^{*} (\bar{S},\bar{C})$.

%s2.2 #&#
\subsection{Model fitting within a Bayesian framework}\label{modfitind}

For data $\{x_{1},x_{2},\ldots,x_{n}\}$, as suggested above, it is easiest
to write the full Bayesian model in terms of the joint distribution $\{
(X_{i},K_{i}), i=1,2,\ldots,n\}$ given $\mu,\sigma^{2}$ with a prior on\vspace*{1pt}
$\mu$ and~$\sigma^{2}$, that is, as $\Pi_{i} \frac{1}{\sigma}
\varphi
((x_{i} + 2\pi k_{i} - \mu)/\sigma)[\mu,\sigma^2]$. Hence, the
posterior for this model involves the latent $\{K_{i}\}$ (only employed
to facilitate model fitting) as well as $\mu$ and $\sigma^{2}$. The
$K_{i}$'s will be updated in implementing a Gibbs sampler,\vspace*{1pt} but only the
posterior samples of $\mu$ (which will provide posterior samples of
$\tilde{\mu}$) and $\sigma^{2}$ will be of interest.
%In fact, we will obtain posterior samples say $(\mu_{b}^{*},
%posterior samples for the concentration, $c$.
To update the $K_{i}$, since we are given $\mu$ and $\sigma^2$, we can
use~\eqref{rule1} to implement \textit{adaptive} truncation, that is, we
can take $m=1+\lfloor\frac{3\sigma}{2\pi}\rfloor$ and $k=\{
-m,\ldots
,0,\ldots,m \}$. Then,
%
%e6 #&#
\begin{eqnarray} \label{threeK}
\operatorname{Pr}(K_{i}=k_{i}|\mu,\sigma,x_{i}) \approx
\frac{\varphi
((x_{i}+2k_{i}\pi- \mu)/\sigma)} {
\sum_{k_{i}=-m}^{m}\varphi((x_{i}+2k_{i}\pi-\mu)/\sigma)}
\nonumber
\\[-8pt]
\\[-8pt]
\eqntext{k_{i}=-m,\ldots,0,\ldots,m.}
\end{eqnarray}
Thus, we achieve constant maximum approximation error at each iteration.

The discussion of the previous section helps in prior specification.
First, as it is customary, we assume $\mu$ and $\sigma^2$ are
independent. For $\mu$ we would adopt a normal distribution, say,
something like $N(\mu_{0}, \sigma_{0}^{2})$. Recalling that $\mu=
\tilde{\mu} + K_{\mu}$, this induces a WN prior on $\tilde{\mu}$ but
also makes it clear that we cannot learn about $\mu$ from the
$X_{i}$'s, that is, in (\ref{threeK}), we cannot identify the
$k_i$'s\vadjust{\goodbreak}
and $k_{\mu}$, hence the $k_{i}$'s and $\mu$. However, we can learn
about $\tilde{\mu}$. Furthermore, due to the conjugacy, we obtain a
familiar normal for the full conditional for $\mu$, that is, $N(\frac
{\sigma_{0}^{2}\sum_{i}(x_{i}+2\pi k_{i}) + \sigma^{2} \mu_{0}}{n\sigma_{0}^{2} + \sigma^2}, \frac{\sigma^{2} \sigma_{0}^{2}}{\sigma^{2} +
n\sigma_{0}^{2}})$. For $\sigma^2$, from the previous section, we
suggest a right truncated inverse Gamma with known scale $\beta_0$ and
shape $\alpha_0$ and truncation defined according to $\hat{\sigma}^{2}$
and $n$, as in Section~\ref{wnd}. For example, if the sample size is $n=30$,
the inverse gamma can be right truncated at $\pi$. Then, the full
conditional for $\sigma^2$ will be a right truncated inverse gamma with
shape parameter $\alpha_0+n/2$ and scale parameter $\beta_0+\frac
{1}{2}[\sum_{i=1}^n (x_i+2k_i\pi-\mu)^{2}]$.
With such priors, the $k_{i}$ are still updated as in (\ref{threeK}).
%The full conditional for $\mu$ will become a wrapped normal with mean $
%full conditional for $\sigma^2$ will be a right truncated inverse
%gamma with shape parameter $\alpha_0+n/2$ and scale parameter $\beta_0+
%for $K_{\mu}$ will be sampled similarly to (\ref{threeK}).

The MCMC is straightforward, though convergence for $\mu$ and the
$K_{i}$'s will not be achievable due to the identifiability problem.
However, we can perform usual convergence diagnostics on the $\mu+2\pi
K_{i}$ on $\tilde{\mu}$ and on $\sigma^2$'s, all of which are well
identified. Then, the posterior samples of $\sigma^2$ inform about
posterior features for~$\sigma^2$, and hence for $c$. The posterior
samples of $\mu$ yield posterior samples of $\tilde{\mu}$; for the
latter, we can adopt whatever posterior centrality summary we wish.
Attractively, we can directly create a $1-\alpha$ credible set, that
is, a symmetric posterior credible arc [Fisher (\citeyear{fisher})]. This is merely
the arc that contains the central $1-\alpha$ proportion of the
posterior samples.

%s3 #&#
\section{Wrapped Gaussian processes}\label{WGP}
Here, we show how a Gaussian process model for linear spatial data
induces a spatial process model for wrapped data. We examine some of
the properties of the induced wrapped process, in particular, the
induced covariance structure. We discuss model fitting for directional
data obtained at a collection of spatial locations using this process.
Again, we adopt a hierarchical modeling approach, describing model
fitting within the Bayesian framework using MCMC. We briefly look at
regression in the context of wrapped periodic data. Finally, we show
how to implement Bayesian kriging, that is, spatial prediction, within
this framework. As with Bayesian kriging in the context of usual
Gaussian processes, we are able to implement such prediction as a
post-model fitting activity.

%s3.1 #&#
\subsection{Multivariate wrapped distributions}\label{MWD}

There is surprisingly little literature on multivariate directional
data modeling. Bivariate circular distributions (whence the support is
a torus) are discussed in \citet{mardiajupp},
\citet{Jamma}. In \citet{kato08}
bivariate circular Cauchy distributions are considered in the
circular--circular regression framework (Section~\ref{regWN}). In this setting,
there is effort to define a sensible measure of correlation between
such pairs and to test for independence for such pairs (Section~\ref{fitWGP}).
We shall see that things simplify when we work with wrapped normal
distributions. Our multivariate motivation is a setting where the
directional data are wave directions at locations and there is
anticipated spatial dependence between the angular variables.
%In our example of Section 5, we work with a set of wave directions
%obtained as computer model output over a grid of cells. \textbf{G $\&$
%D: Why anticipate here the case study?}

As a general strategy, it is perhaps easiest to obtain a multivariate
wrapped distribution for, say, $\mathbf{X} = (X_{1},X_{2},\ldots,X_{n})$
starting with a multivariate linear distribution for $\mathbf{Y} =
(Y_{1},Y_{2},\ldots,Y_{n})$. In particular, suppose $\mathbf{Y} \sim
g(\cdot)$, where $g(\cdot)$ is a $n$-variate distribution on $\mathbb
{R}^{n}$. Usually, $g$ is a family of distributions indexed by, say,
$\btheta$; a convenient choice for $g(\cdot)$ is an $n$-variate normal
distribution. Let $\mathbf{K} = (K_{1},K_{2},\ldots,K_{n})$ be such that
$\mathbf{Y}= \mathbf{X} + 2\pi\mathbf{K}$, analogous to the univariate
case. Then, the joint distribution of $\mathbf{X}$ and $\mathbf{K}$ is
$g(\mathbf{x}+ 2 \pi\mathbf{k})$ for $0 \leq x_{j} < 2 \pi,
j=1,2,\ldots,n$ and $k_{j} \in\mathbb{Z}, j=1,2,\ldots,n$. The marginal
distribution of $\mathbf{X}$ is an $n$-fold doubly infinite sum of
$g(\mathbf{x}+ 2 \pi\mathbf{k})$ over $\mathbb{Z}^{n}$.
Its form is intractable to work with, even for moderate $p$. Again, we
introduce latent $K_{j}$'s to facilitate the model fitting [see \citet
{coles98} in this regard and Section~\ref{fitWGP} below].

We say that $\mathbf{X}$ has a $p$-variate wrapped normal distribution
when $g(\cdot; \btheta)$ is a multivariate normal where $\btheta=
(\bmu
, \bSigma)$, with $\bmu$ an $n \times1$ vector of mean directions and
$\Sigma$ a positive definite matrix. Using standard results, the
conditional distribution of $Y_{j}$ given $\{Y_{l}, l \neq j\}$ and
$\btheta$ is immediate, hence, as well, the distribution of
$X_{j},K_{j}$ given $\{X_{l},K_{l}, l \neq j \}$ and $\btheta$.

%s3.2 #&#
\subsection{Wrapped spatial Gaussian processes}\label{WSGP}

A Gaussian process on $\mathbb{R}^{d}$ induces a wrapped Gaussian
process on $\mathbb{R}^{d}$. In particular, the Gaussian process (GP)
is specified through its finite dimensional distributions which in turn
induce the finite dimensional distributions for the wrapped process.
Hence, we are returned to the multivariate wrapped distributional
models of the previous subsection. In particular, if $s \in\mathbb
{R}^{d}$ and $Y(s)$ is a GP with mean $\mu(s)$ and covariance function,
say, $\sigma^{2} \rho(s-s'; \phi)$, where $\phi$ is a decay parameter,
then, for locations $s_{1},s_{2},\ldots,s_{n}$, $\mathbf
{X}=(X(s_{1}),X(s_{2}),\ldots,X(s_{n})) \sim \operatorname{WN}(\bolds{\mu
},\sigma^{2}\mathbf{R}(\phi))$, where $\bolds{\mu}= (\mu
(s_{1}),\ldots, \mu
(s_{n}))$ and $R(\phi)_{ij}= \rho(s_{i} - s_{j};\phi)$. In the sequel
we utilize a stationary, in fact, isotropic covariance function but
with regard to the model fitting (see below); other choices could be
investigated similarly. We note that the multivariate wrapped modeling
in \citet{coles98} employs replications to learn about a general
$\bolds
{\Sigma}$ for the multivariate model. We do not require replications
due to the structured spatial dependence introduced through the GP.
Also, to implement a spatial regression model for angular response with
linear covariates, it is necessary to introduce a monotone link
function $h(\cdot)$ from $R^1$ to $(-\pi, \pi)$ with $h(0)=0$, for
example, $h(\cdot) = \operatorname{arctan}(\cdot)$ [\citet{Lee2010}].
%(Lee, 2010 \textbf{G - Add to ref list, Lee, A. (2010), Circular Data,
%Wiley Interdisciplinary Reviews: Computational Statistics, 2, 477-486}.
In the sequel, we confine ourselves to the case where $\mu(s) = \mu$.
%whence the prior specification must include a mean direction. Again,
%our primary objective is to incorporate spatial dependence into the $

%s3.3 #&#
\subsection{Fitting a wrapped GP model}\label{fitWGP}

Model fitting for a wrapped GP within a Bayesian framework can be done
using MCMC. First, suppose a linear GP model of the form $Y(s_{i}) =
\mu+ w(s_{i}), i=1,2,\ldots,n$, where $w(s_{i})$ is a mean $0$ GP with
covariance function $\sigma^{2}\rho(s - s';\phi)$. Consider an
exponential covariance and a prior on $\btheta= (\mu, \sigma^{2},
\phi
)$ of the form $[\mu][ \sigma^2][ \phi]$ which is normal, inverse Gamma
and uniform, respectively. Because of the well-known identifiability
issue with $\sigma^2$ and $\phi$ [\citet{zhang04}], it is often best to
update them as a pair on the log scale using a Metropolis--Hastings step
with a bivariate normal proposal, following the implementation scheme
underlying the R package spBayes [\citet{spBayes}]. For the wrapped GP,
the approach follows that of Section~\ref{univ} by introducing a latent vector
of $K$'s. Again, the induced wrapped GP provides $\mathbf{X} \sim
\operatorname{WN}(\mu\mathbf{1}, \sigma^{2}\mathbf{R}(\phi))$, where $R(\phi
)_{jk} =
\rho(s_{j}-s_{k};\phi)$ so that the joint distribution of $\mathbf{X},
\mathbf{K}$ takes the form $N(\mathbf{X}+2\pi\mathbf{K} |\mu
\mathbf
{1}, \sigma^{2}\mathbf{R}(\phi))$. From above, we suggest a normal
prior for $\mu$, a truncated inverse gamma prior for $\sigma^2$ and,
for the decay parameter, we have employed a uniform prior with support
allowing small ranges up to ranges a bit larger than the maximum
distance over the region. The full conditionals for~$\mu$ and $\sigma^2$
are similar to those in Section~\ref{univ}. The full conditional for~$\phi$
is unpleasant because it is buried in the covariance matrix associated
with the $n$-variate normal distribution for $\mathbf{X}+2\pi\mathbf
{K}$. Again, we have found it best to update~$\sigma^2$ and $\phi$ as a
pair on the log scale as in the linear case. In fact, with such joint
sampling, very large values of $\sigma^2$ are rejected in the M--H step,
so, in fact, we do not need to impose any truncation on the prior for
$\sigma^2$.

Finally, the full conditionals for the $K_{i}$ arise from the
conditional distribution of $Y_{i}=X_{i}+2\pi k_{i}|\{Y_{j} =
X_{j}+2\pi K_{j}, j \neq i\}; \btheta$. The form is analogous to (\ref
{threeK}) with $\mu$ and $\sigma^2$ replaced by the conditional mean
and variance $\mu_{i}$ and $\sigma_{i}^{2}$ which are functions of $\{
X_{j}, j \neq i \}$, $\{K_{j}, j \neq i \}$ and $\theta$. The adaptive
approximation of Section~\ref{wnd} can be employed.

Moments estimates associated with the wrapped Gaussian process are
useful for the same two purposes as in the independence setting. One is
to help specify priors to facilitate inference stability, following the
discussion of Section~\ref{univ}. The other is to provide a sensible range for
starting values to begin the MCMC model fitting.
Under normality, again, $E(e^{iX(s)})= \operatorname{exp}(i\mu-\sigma^{2}/2)$.
So, moment estimators for mean direction, uncertainty and concentration
are as in the independence case.

%s3.4 #&#
\subsection{Induced correlation for the circular variables}\label{correlation}

%It is useful to investigate first and second moment properties
%associated with the wrapped Gaussian process for the same two purposes
%as in the univariate setting. One is to introduce priors on the model
%parameters to facilitate inference stability, following the discussion
%of Section 2. The other is to provide a sensible range for starting
%values to begin the MCMC that we use for the model fitting.
%With regard to moments, again we work with the associated complex
%variables on the unit circle, $e^{iX(s)} = \operatorname{cos}(X(s)) + i
%$E(e^{ipX(s)})= E(e^{ipY(s)})$ for all positive integers $p$.
%In particular, under normality, again, $E(e^{iX(s)})=
%= e^{-\sigma^{2}/2}$ are as in the univariate case.
%Also, $E(e^{2iX(s)})= \operatorname{exp}(2i\mu-2\sigma^{2})$ and, since
%$Y(s_{1})+Y(s_{2}) \sim N(2\mu, 2\sigma^{2}(1+\rho(s_{1}-s_{2};
%)})= \operatorname{exp}(2i\mu- \sigma^{2}(1+\rho(s_{1}-s_{2};\phi)))$.
%These
%calculations yield expressions for ``variances'' and ``covariances''
%though, as complex numbers, they are not immediately useful.

\begin{figure}

\includegraphics{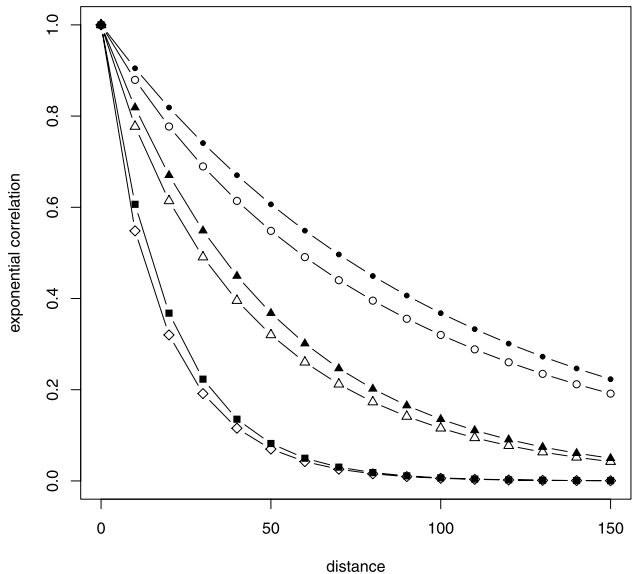}

\caption{Exponential spatial correlation function (solid circle $\phi
=0.01$, solid triangle $\phi=0.02$, solid square $\phi=0.05$) and the
corresponding wrapped correlation (empty circle $\phi=0.01$, empty
triangle $\phi=0.02$, empty square $\phi=0.05$) for 3 values of the
decay parameter $\phi$.}
\label{corcirc}
\end{figure}

We have defined the wrapped GP in terms of the linear GP which has a
covariance or correlation structure (and the parameters which specify
them). For a bivariate circular variable there is no unique choice of
correlation measure. However, in \citeauthor{Jamma} [(\citeyear{Jamma}), Chapter 8], we find
considerable discussion of suitable measures for the correlation
between two circular variables. \citet{jamma88} propose a measure which
satisfies many of the properties of the product moment correlation
coefficient. For the wrapped bivariate normal for variables $X_{1},
X_{2}$ with covariance matrix $\bigl(
{
{\sigma^{2} \atop \rho\sigma^{2}}\enskip
{\rho\sigma^{2} \atop \sigma^{2}}}
\bigr)$
it simplifies to $\rho_{c}(X_{1},X_{2}) = \operatorname{sinh}(\rho
\sigma^{2})/\operatorname{sinh}(\sigma^{2})$. Hence, with a valid
covariance function
$\sigma^{2}\rho(s,s')$, the induced correlation function for the
wrapped Gaussian process is $\rho_{c}(s,s')= \frac{\operatorname
{sinh}(\sigma^{2}\rho(s,s'))}{\operatorname{sinh}(\sigma^{2})}$. Figure~\ref{corcirc}
provides a picture of the exponential correlation function for various
choices of the decay parameter~$\phi$ and the corresponding correlation
for the wrapped process. For a given distance, association is similar
but a bit weaker under the latter.

%f1 #&#

As an aside, though we don't build any covariance matrices using
$\operatorname
{sinh}(\rho(u))$, we note that it is a valid covariance function on
$\mathbb{R}^{d}$ if $\sigma^{2} \rho(u)$ is. This is evident since
$\operatorname{sinh}(v) = \sum_{n=1}^{\infty} \frac{v^{2n+1}}{(2n+1)!}$.
Because $\rho(u)$ is a valid correlation function, hence a
characteristic function by Bochner's theorem, $\rho^{p}(u)$ is a
characteristic function for any integer $p>0$. So, $\operatorname
{sinh}(\rho
(u))$ is a mixture (with suitable normalization) of characteristic
functions, hence a characteristic function, therefore, again, by
Bochner's theorem, a valid covariance function itself.

%s3.5 #&#
\subsection{Circular--circular regression with WN models}\label{regWN}

For general bivariate directional data models, say, $g(X_{1},X_{2})$,
circular--circular regression for, say, $X_{1}$ given $X_{2}$, is
discussed in \citet{jamma93} using trigonometric polynomial
approximations. In the case of wrapped bivariate normal distributions,
we can obtain the regression explicitly, avoiding approximation. We
also note that Kato, Shimizu and Shieh (\citeyear{kato08}) consider circular--circular regression
curves obtained under\vadjust{\goodbreak} a M\"{o}bius circle transformation [See also
\citet{downs-mardia02} in this regard].
Incidentally, they discuss a bivariate circular distribution which is
specified through the product of a conditional times a marginal wrapped
Cauchy distribution.

We return to the wrapped bivariate normal for variables $X_{1}, X_{2}$
with mean $\mu_{1}, \mu_{2}$ and covariance matrix $\bigl(
{
{\sigma^{2} \atop \rho\sigma^{2} }\enskip
{\rho\sigma^{2} \atop \sigma^{2}}}
\bigr).$
The regression model we seek is $E(e^{iX_{2}}|X_{1}; \btheta)$ in the
form $c(X_{1};\btheta) e^{i \mu(X_{1};\btheta)}$, where $\btheta
=(\mu_{1},\mu_{2}, \sigma^2, \rho)$. Then, we can interpret, respectively,
$\mu(X_{1};\btheta)$ as the conditional mean direction and
$c(X_{1};\btheta)$ as the conditional concentration for $X_{2}$ given $X_{1}$.

For the associated unwrapped normal distribution,
\[
Y_{2}|Y_{1};\btheta\sim N\bigl(\mu_{2|1}(Y_{1};
\btheta), \sigma^{2}_{2|1}(\btheta)\bigr),
\]
where $\mu_{2|1}(Y_{1};\btheta) = \mu_{2}+\rho(Y_{1}-\mu_{1})$ and
$\sigma^{2}_{2|1}(\btheta)= \sigma^{2}(1-\rho^{2})$. Hence, writing
$Y_{1}= X_{1} + 2\pi K_{1}$, we have $X_{2}|X_{1}, K_{1};\btheta\sim
\operatorname{WN}(\tilde{\mu}_{2|1}(X_{1}+ 2 \pi K_{1});\btheta), \sigma^{2}_{2|1}(\btheta)),$ where $\tilde{\mu}(Y)= \mu(Y) \operatorname
{mod}2 \pi$
and, therefore,
\[
E\bigl(e^{iX_{2}}|X_{1},K_{1};\btheta\bigr) =
\operatorname{exp}\bigl(-\sigma^{2}_{2|1}(\btheta )/2 + i
\tilde{\mu}_{2|1}(X_{1}+2\pi K_{1};\btheta)\bigr).
\]

Next, we have $E(e^{iX_{2}}|X_{1};\btheta)= E_{K_{1}|X_{1};\btheta
}E(e^{iX_{2}}|X_{1},K_{1};\btheta),$ where the conditional distribution
for $K|X, \btheta$ under the wrapped normal is discussed in Section~\ref{univ}.
So, let $p(k;X_{1},\btheta) = P(K_{1}=k|X_{1},\btheta)$. Then,
%
%e7 #&#
\begin{eqnarray}
E\bigl(e^{iX_{2}}|X_{1};\btheta\bigr)&=& \operatorname{exp}
\bigl(-\sigma^{2}_{2|1}(\btheta )/2\bigr)
\nonumber
 \\[-8pt]
 \\[-8pt]
 \nonumber
&&{}\times \sum_{k=-\infty}^{\infty}
p(k;X_{1},\btheta) \operatorname {exp}\bigl(i \tilde{\mu
}_{2|1}(X_{1}+2\pi k;\btheta)\bigr).
\end{eqnarray}
We see that $c(X_{1};\btheta)= e^{-\sigma^{2}_{2|1}(\btheta)/2}$ and
$\operatorname{exp}(i \mu(X_{1};\btheta))= \sum_{k=-\infty
}^{\infty}
p(k;X_{1},\btheta) \times \operatorname{exp}(i \tilde{\mu
}_{2|1}(X_{1}+2\pi k;\btheta
))$. So, $\operatorname{cos}(\mu(X_{1};\btheta)) = \sum_{k=-\infty
}^{\infty}
p(k;X_{1},\btheta) \operatorname{cos}( \tilde{\mu}_{2|1}(X_{1}+2\pi
k;\btheta
))$ and $\operatorname{sin}(\mu(X_{1};\btheta)) = \sum_{k=-\infty
}^{\infty}
p(k;X_{1},\btheta) \operatorname{sin}( \tilde{\mu}_{2|1}(X_{1}+2\pi
k;\btheta
))$. Making the usual inversion,
%
%e8 #&#
\begin{equation}
\mu(X_{1};\btheta) \equiv\operatorname{arctan}^{*} \bigl(
\operatorname{sin}\bigl(\mu (X_{1};\btheta)\bigr),\operatorname{cos}
\bigl(\mu(X_{1};\btheta)\bigr) \bigr).
\end{equation}
In practice, we would compute $\mu(X_{1};\btheta)$ by appropriate
truncation of $K$. If we fit the bivariate wrapped normal distribution
with data $(X_{1i},X_{2i}), i=1,2,\ldots,n$, using MCMC, posterior samples
for $\btheta$ enable posterior samples for $\mu(X_{1};\btheta)$ and
$c(X_{1};\btheta)$ at any $X_{1}$.

%s3.6 #&#
\subsection{Kriging with wrapped GP models}\label{krigtheory}

Kriging is a customary activity with spatial data. In this context, we
would have, as observations, $\mathbf{X} =
(X(s_{1}),X(s_{2}), \ldots,X(s_{p}))$ and we would seek to predict
$X(s_{0})$ at a new location $s_{0}$. In fact, we shall argue that this
is a straightforward post-model fitting exercise and can be implemented
following the ideas of circular--circular regression for the wrapped
normal from the previous subsection.

Suppose, for the linear ``observations,'' $\mathbf{Y} =
(Y(s_{1}),Y(s_{2}),\ldots,Y(s_{p}))$ along with $Y(s_{0})$ we have the
joint distribution
%
%e9 #&#
\begin{equation}
\pmatrix{ \mathbf{Y}
\vspace*{2pt}\cr
Y(s_{0})}
= N\biggl( \pmatrix{
 \bmu
\vspace*{2pt}\cr
\mu(s_{0})}, \sigma^{2} \pmatrix{\mathbf{R}_{\mathbf{Y}}(\phi) &\brho_{0,\mathbf{Y}}(\phi)
\vspace*{2pt}\cr
\brho^{T}_{0,\mathbf{Y}}(\phi) & 1} \biggr).
\end{equation}
Evidently, we can obtain the distribution for $Y(s_{0})|\mathbf{Y},
\btheta$, hence the wrapped normal distribution for $X(s_{0})|\mathbf
{X}, \mathbf{K}, \btheta$ and, thus, $E(e^{i X(s_{0})}|\mathbf{X},
\mathbf{K}, \btheta)$, following the previous section. Now, for $E(e^{i
X(s_{0})}|\mathbf{X}, \btheta)$ we would need to marginalize over the
distribution of $\mathbf{K}|\mathbf{X}, \btheta$. This requires a
$p$-fold sum over a multivariate discrete distribution, hopeless for
large $p$ even with considerable truncation, for example, a sum over~$3^{p}$ terms if each $K_{i}$ is allowed only $3$ values.

In fact, within the Bayesian modeling framework we seek $E(e^{i
X(s_{0})}|\mathbf{X})$. Fitting the spatial wrapped GP model as in
Section~\ref{correlation} will yield posterior samples, say, $(\btheta_{b}^{*},\mathbf
{K}_{b}^{*}), b=1,2,\ldots,B$. Then, as usual, $E(e^{i
X(s_{0})}|\mathbf
{X}) = E_{\mathbf{K},\btheta|\mathbf{X}}E(e^{i X(s_{0})}|\mathbf{X},
\mathbf{K}, \btheta)$ and so a Monte Carlo integration yields
%
%e10 #&#
\begin{equation}
\quad E\bigl(e^{i X(s_{0})}|\mathbf{X}\bigr) \approx\frac{1}{B} \sum
_{b} \operatorname {exp}\bigl(-\sigma^{2}
\bigl(s_{0}, \btheta_{b}^{*}\bigr)/2+ i \tilde{
\mu}\bigl(s_{0}, \mathbf{X}+2\pi\mathbf{K}_{b}^{*};
\btheta_{b}^{*}\bigr)\bigr),
\end{equation}
where $\tilde{\mu}(s_{0}, \mathbf{Y}, \btheta)$ and $\sigma^{2}(s_{0},
\btheta)$ extend the notation $\tilde{\mu}_{2|1}$ and $\sigma_{2|1}^{2}$ of the previous section to $Y(s_{0})|\mathbf{Y}$. If
$g_{c}(s_{0}, \mathbf{X}) = \frac{1}{B} \sum_{b^{*}} \operatorname
{exp}(-\sigma^{2}(s_{0}, \btheta_{b}^{*})/2)\operatorname{cos}( \tilde{\mu
}(s_{0}, \mathbf
{X}+2\pi\mathbf{K}_{b}^{*};\btheta_{b}^{*}))$ and $g_{s}(s_{0},
\mathbf
{X}) = \frac{1}{B} \sum_{b^{*}} \operatorname{exp}(-\sigma^{2}(s_{0}, \btheta_{b}^{*})/2)\operatorname{sin}( \tilde{\mu}(s_{0}, \mathbf{X}+2\pi
\mathbf
{K}_{b}^{*};\btheta_{b}^{*}))$, then the posterior mean kriged
direction is
%
%e11 #&#
\begin{equation}
\label{krigmc} \mu(s_{0}, \mathbf{X}) = \operatorname{arctan}^{*}
\bigl(g_{0,s}(\mathbf {X}),g_{0,c}(\mathbf{X}) \bigr)
\end{equation}
and the associated posterior kriged concentration is
%
%e12 #&#
\begin{equation}
\label{postconcentration} c(s_{0}, \mathbf{X}) = \sqrt{
\bigl(g_{c}(s_{0}, \mathbf {X})\bigr)^{2}+
\bigl(g_{s}(s_{0}, \mathbf{X})\bigr)^{2}}.
\end{equation}

%s4 #&#
\section{Examples and MCMC implementation}\label{examples}
In Section~\ref{sectionsim} we offer some simulation examples, while in
Section~\ref{sectionrd} we turn to the motivating spatial wave
direction data. We fit the model described in Section~\ref{WSGP} and
implemented the kriging following Section~\ref{krigtheory}. We note that MCMC model
fitting as described in Section~\ref{correlation} is well behaved.

%s4.1 #&#
\subsection{Simulation examples}\label{sectionsim}

In the simulations we generate samples of size $n=100$ from an
unwrapped GP with constant mean $\mu$ and we obtain the directional
process by wrapping them onto the unit circle. Locations are generated
uniformly using coordinates taken from the real data example of Section
\ref{sectionrd}; we work with two different sample sizes by randomly
choosing 30 and 70 sites for posterior estimation and the remaining are
used for validation.\vadjust{\goodbreak} We fix the covariance structure for the linear GP
to be exponential with parameters $(\sigma^2, \phi)$. Here we report
examples generated with $\phi=0.021$ corresponding to a practical range
of 142.86~km (maximum distance spanned by the coordinates is 290.13~km), $\mu=\pi$ and three different variances $\sigma^2=0.1,0.5,1$
corresponding to concentrations $c=0.951,0.779,0.606$ respectively. The
prior for $\mu$ is a Gaussian distribution with zero mean and large
variance (with an induced wrapped normal prior for~$\tilde{\mu}$). For
$\sigma^2$ we use informative inverse gamma distributions centered on
the \textit{true} value and variances 0.01, 0.06 and 0.07, respectively.
For the decay parameter the prior is a uniform distribution in $[0.001,
1)$ when $\sigma^2=0.1, 0.5$ and in $[0.001,0.5)$ when $\sigma^2=1$.
For all priors settings, several variance values were considered to
assess behavior under strongly and weakly informative priors. The block
sampling of $\sigma^2$ and $\phi$ produces strongly autocorrelated
chains for both parameters and slow convergence compared to the
independence case. We run the MCMC for 30,000 iterations, we discard the
initial 6000 and we apply a thinning of 10, using 2400 samples for estimation.

The precision of estimates is studied by $95\%$ credible intervals,
obtained as discussed in Section~\ref{modfitind}. We compute kriging estimates as
in \eqref{krigmc} and \eqref{postconcentration} for each specification
and compute an average prediction error, defined as the average
circular distance\footnote{We adopt as circular distance, $d(\alpha
,\beta)=1-\cos(\alpha-\beta)$, as suggested in Jammalamadaka and
SenGupta [(\citeyear{Jamma}), page 16].} between the observed and kriged estimate,
over the validation set of observed values. That is,\vspace*{-1pt} for the validation
set $\{x(s_{j}^{*}), j=1,2,\ldots,m\}$, we compute $\frac{1}{m}\sum_{l}(1-\operatorname{cos}(\mu(s_{j}, \mathbf{X}) - x(s_{j}^{*})))$.
We also compare this spatial interpolation to prediction obtained by
fitting the independent wrapped normal model proposed in Section \ref
{univ} and the prior structure described there. Comparison is assessed
through average error, computed for the nonspatial model as the average
circular distance between the directions in the validation set and the
estimated mean direction of the nonspatial model.

In Table~\ref{posterior1} posterior estimates and interpolation errors
are given for both the spatial and nonspatial models.
%Using the posterior samples, point estimates are obtained as posterior
%mean for the mean and the concentration, while the posterior
%distribution modal value is used for the decay parameter being the
%posterior asymmetric (see for example Figure~\ref{priorpostrd} in the
%next section).
Both models can recover the mean direction and concentration with wider
confidence intervals when $c$ is small. The decay parameter is
correctly estimated when $c=0.951,0.779$ for both sample sizes but
requires larger sample size when $c=0.606$.
There is substantial reduction in average prediction error using the
wrapped GP model when there is spatial dependence, while for small
ranges (not shown) the spatial model performs comparably to the
nonspatial one.
%

%t1 #&#
\begin{table}
\caption{Results from simulated data with $\mu=\pi$ and $\phi=0.013$
for all simulations. Posterior mean and concentration point estimates
are obtained averaging the MCMC samples, while the decay point estimate
is obtained as modal value}\label{posterior1}
\begin{tabular*}{\textwidth}{@{\extracolsep{\fill}}lcccc@{}}
\hline
& $\bolds{\hat{\mu}}$ & $\bolds{\hat{c}}$ & $\bolds{\hat{\phi
}}$ & \textbf{Average} \\
& \textbf{(95\% CI)} &\textbf{(95\% CI)} & \textbf{(95\% CI)}&
\textbf{prediction error} \\
\hline
$c=0.951$& &&&\\
$n=30$ & 3.124 & 0.926 & 0.013 & 0.034 \\
& {$(2.716 , 3.517)$} & {$(0.825,0.965)$} &
{$(0.005,0.066)$}& \\
$n=70$ & 3.135 & 0.929 & 0.017 & 0.023 \\
& {$(2.800 ,3.505)$} & {$(0.848 ,0.962)$} & {$(0.007 ,
0.037)$}& \\[3pt]
$c=0.779$&&&&\\
$n=30$ &3.164 & 0.708& 0.018 & 0.058 \\
& {$(2.399 , 3.777)$} & {$(0.547, 0.811)$} & {$(0.009 ,
0.329)$}& \\
$n=70$ & 3.229 & 0.748 & 0.015 & 0.085 \\
& {$(2.518,3.898)$} & {$(0.596,0.841)$} &
{$(0.008,0.032)$}& \\[3pt]
$c=0.606$&&&&\\
$n=30$ & 2.916 & 0.594 & 0.049 & 0.188 \\
& {$(2.416 ,3.514)$} & {$(0.470, 0.693)$} & {$(0.022,
0.190)$}& \\
$n=70$ & 2.928 & 0.608 & 0.025 & 0.099 \\
& {$(2.261, 3.673)$} & {$(0.480,0.706)$} & {$(0.015,
0.048)$}& \\[6pt]
Non spatial model& &&&\\
$c=0.951$&&&&\\
$n=30$ & 3.102 & 0.947 & & 0.048 \\
&{$( 2.981, 3.223)$} & {$(0.914, 0.968)$}&& \\
$n=70$ & 3.110 & 0.948 & & 0.042 \\
& {$(3.034,3.188)$} & {$(0.928, 0.962)$}&& \\[3pt]
$c=0.779$&&&&\\
$n=30$ & 2.781 & 0.794 & & 0.240 \\
& {$(2.535,3.030)$} & {$(0.680,0.871)$}&& \\
$n=70$ & 2.925 & 0.749 & & 0.170 \\
& {$(2.710,3.140)$} & {$(0.649,0.823)$}&& \\[3pt]
$c=0.606$&&&&\\
$n=30$ & 2.869 & 0.640 & & 0.335 \\
& {$(2.514, 3.217)$} &{$(0.473, 0.765)$}&& \\
$n=70$ & 2.785 & 0.677 & & 0.382 \\
& {$(2.577, 3.001)$} & {$(0.578, 0.755)$}&& \\
\hline
\end{tabular*}
\end{table}

%s4.2 #&#
\subsection{Wave direction data analysis}\label{sectionrd}

The majority of studies carried out on marine data are based on outputs
from deterministic models, usually climatic forecasts computed at
several spatial and temporal resolutions. Since the $1980$s,
deterministic models have been used for weather forecasting with
increasing\vadjust{\goodbreak} reliability. Moreover, in the last decade, sea surface wind
data projections, produced by meteorological models, have been found
accurate enough to be taken as the basis for operational marine
forecasts. Wave heights and outgoing wave directions, the latter being
angular data measured in degrees, are the main outputs of marine
forecasts. The principal provider for global numerical wave forecasts
in Europe is the European Center for Medium-Range Weather Forecasts
(ECMWF), which runs at global medium range (3--5 up to 10 day forecasts,
55~km spatial resolution) and at high resolution short term (3 days, 28
km resolution WAve Model, WAM) models in the Mediterranean Area. WAM
outputs are given in deep waters (more than 100~m depth) on a grid with
about $25\times25$~km cells.

Calibration with buoy data for wave heights was studied [\citet
{jonaetal07}] using a multistep model that included outgoing wave
directions as a categorical variable (categorized according to an 8
sector windrose) not allowing for a full exploitation of the available
information. The main use of these calibrated data is in starting of
runs of shallow water forecast models [SWAN model, see, e.g., \citet
{swanbook}]. The output of these models provide the basis for coastal
alarm in severe weather and for the evaluation of coastal erosion.
These models are built on grids with about 10 km spatial resolution and
they include as inputs calibrated and downscaled wave heights,
bathymetry, wind speed and direction all given at the same spatial
resolution. Wave directions are aligned to the finer grid simply by
taking the circular mean of intersected WAM cells. Here we propose a
fully model-based solution using a downscaling procedure.

\begin{figure}

\includegraphics{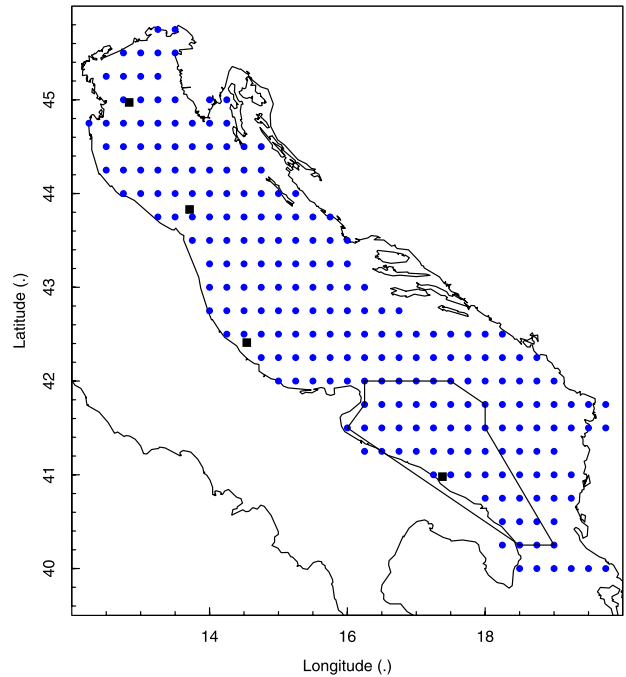}

\caption{WAM grid, Italian Wave Monitoring network in the Adriatic Sea
and estimation area, squares denote locations of buoys.}
\label{ron}
\end{figure}

\begin{figure}

\includegraphics{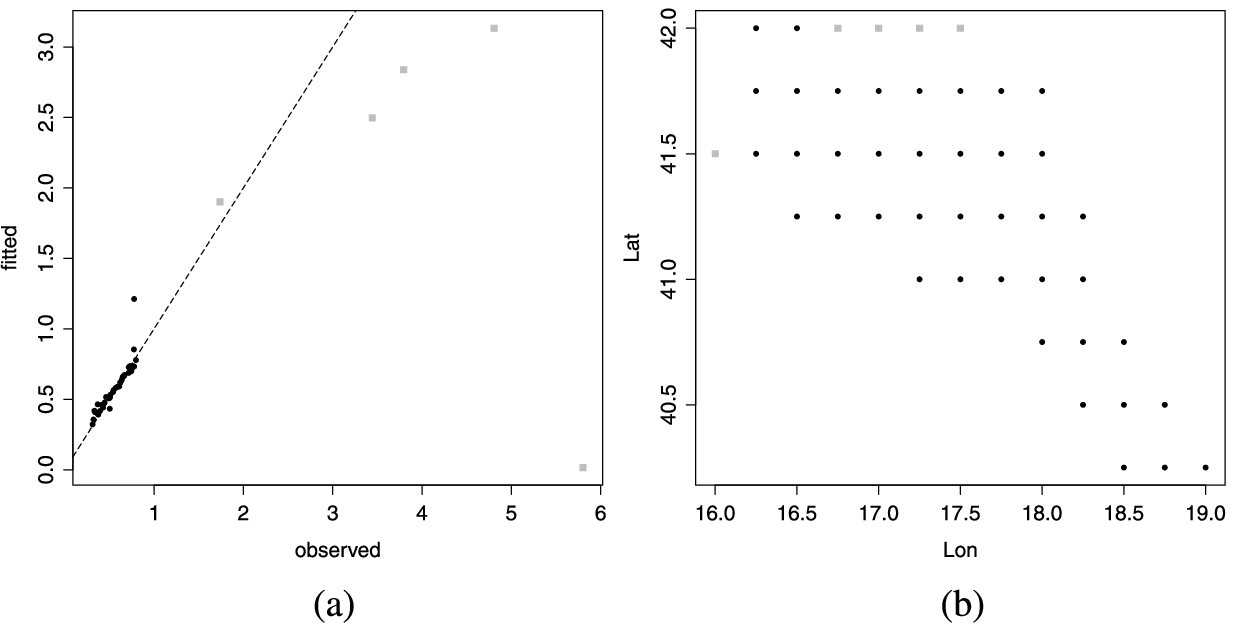}

\caption{\textup{(a)} Kriged estimate vs. observed using leave-one-out
validation, \textup{(b)} locations of the 5~outliers present in the
data (grey squares).}
\label{datapres}
\end{figure}

We analyze data from a single time during a storm in the Adriatic Sea.
The data are outgoing wave directions produced by the WAM during the
analysis, that is, a~run of the forecast of the model at time $t=0$. We
use 45 points from the WAM grid, as they cover a fairly homogeneous
area during the storm movement. In Figure~\ref{ron} the entire Adriatic
network of buoys is shown together with the WAM grid and the estimation
area (delimited by continuous lines). Available data yield a moments
estimated sample concentration of $\hat{c}=0.8447$, in the North--East
direction (moments estimated sample mean direction $\hat{\tilde{\mu
}}=0.5540$). In the data only five values differ from $\pi/4$ by more
than 0.8 radians [they are marked on Figure~\ref{datapres}(b)], four are located in
the upper north part of the area outside the Gargano Peninsula, one is
located near the coast inside the Mattinata Gulf. This last value,
being located in the curve of the gulf, may account for a local wave
direction inversion. The other four seem to describe some turbulence in
the wave field. We keep them in the data set to see how the spatial
model deals with this local behavior. We are interested in downscaling
these \textit{WAM} values to a grid of 222 cells with 10~km resolution.

%f2 #&#

%f3 #&#
\begin{figure}

\includegraphics{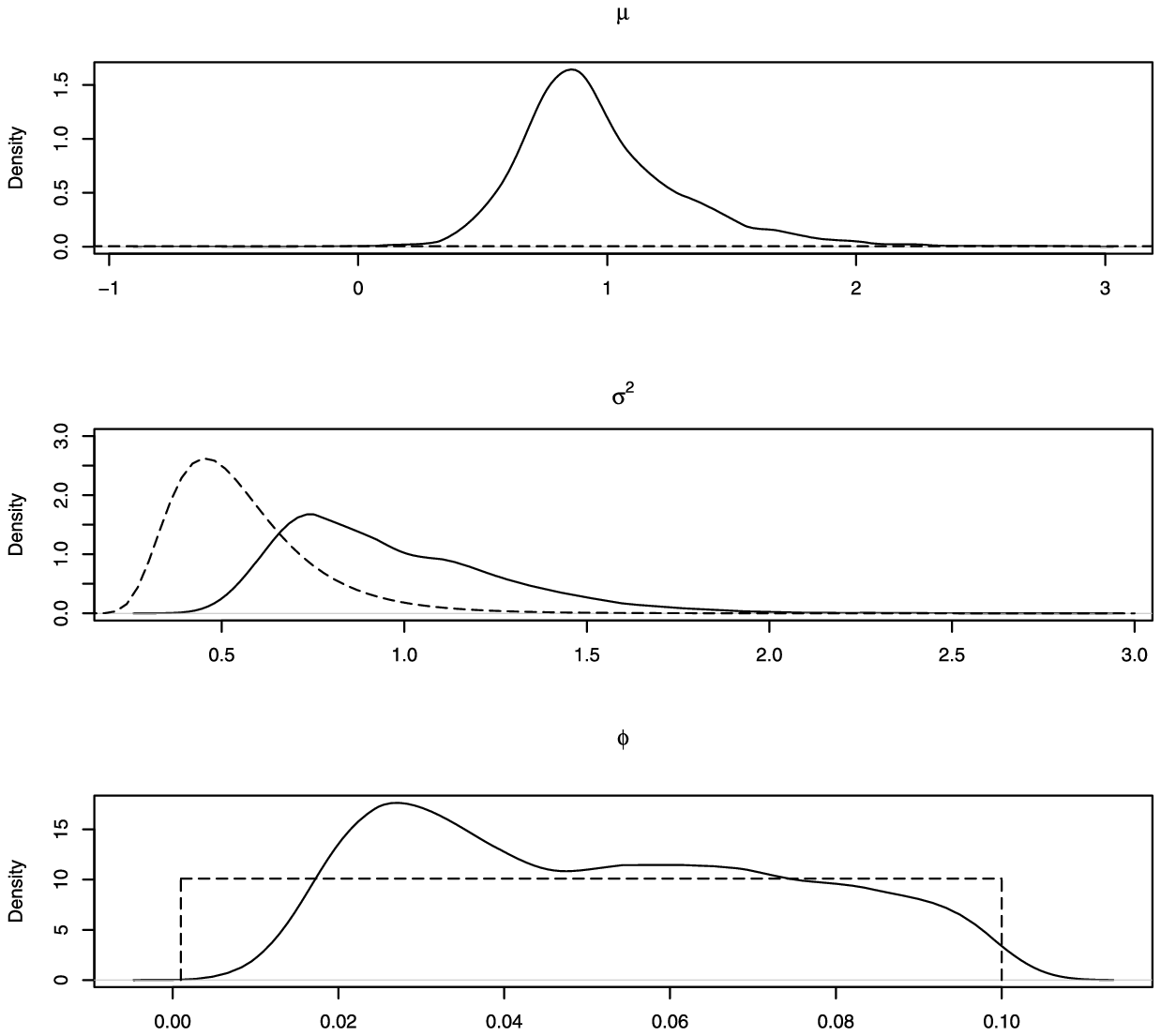}

\caption{Prior (dashed) and posterior (solid) distributions for the WAM
data.}\label{priorpostrd}
\end{figure}

Arguably, an angular data model such as a multivariate
version of the von Mises distribution might be more natural here than a
periodic data model like our wrapped normal. However, because the WN
allows convenient extension to the spatial setting and because there is
little practical difference between the WN and von Mises on the circle,
we work with the latter, as detailed in Section~\ref{WGP}. In the
absence of covariates, we employ the constant mean direction model
(though a trend surface could be investigated). In fitting, we run the
MCMC algorithm for 30,000 iterations, with a burn-in of 6000, and
compute kriging estimates as in \eqref{krigmc} and \eqref
{postconcentration} using 2400, taking one sample every 10 samples. We
choose the following prior setting: $\sigma^2\sim \operatorname{IG}(9,4)$ (i.e.,
mode${}=0.4$ and variance${}=0.04$), $\phi\sim \operatorname{Unif}(0.001,0.1)$ and, for
the mean, a Normal distribution with zero mean and large variance. In
Figure~\ref{priorpostrd} the posterior learning from the data is shown.
Posterior estimates return a posterior mean direction of 0.971 radians
($[0.471 , 1.814]$ is a 95$\%$ credible interval) and posterior
concentration of 0.618 ($[ 0.427,0.753]$ is 95$\%$ credible interval).
The decay parameter, estimated as posterior modal value, is 0.023
($[0.017,0.096]$ is a 95$\%$ credible interval).
In Figure~\ref{datapres} results from a leave-one-out validation
procedure are shown. It can be seen that the five outliers are shrunk
toward the $45^{\circ}$ line. The average prediction error is 0.0488.
%%Sensitivity analysis reveals that average prediction error is robust
%to prior specifications.

%f4 #&#

%f5 #&#
\begin{figure}[b]

\includegraphics{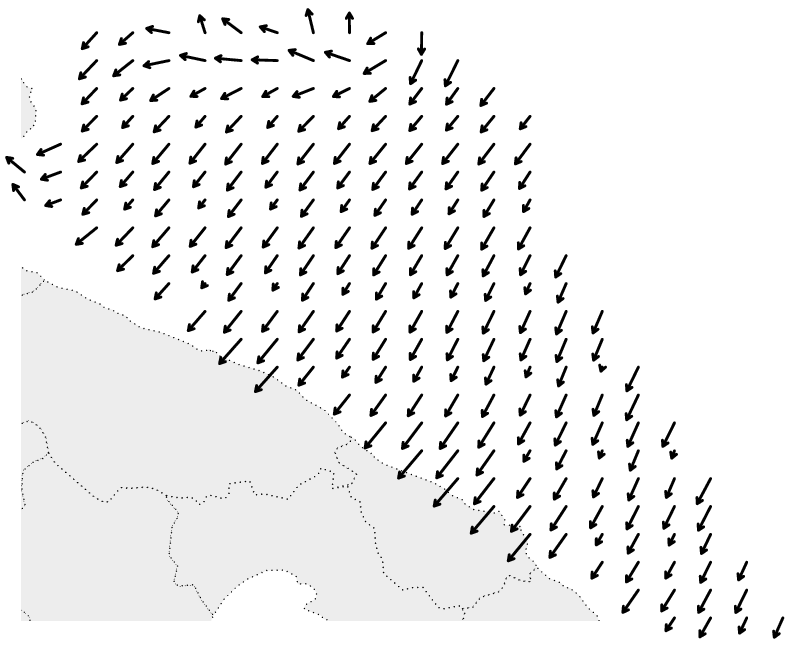}

\caption{Kriging results for WAM data: wave directions represented as
arrow, the length of the arrows is
proportional to 1-concetration. Here we report incoming waves
directions. }\label{krigrd}
\end{figure}

In Figure~\ref{krigrd} an arrow plot of incoming wave directions with
length of the arrows proportional to $1-\hat{c}$ (the longer the
vector, the more variable the estimates at that grid point) is shown.
In the arrow plot we choose to present the incoming direction\footnote
{Incoming wave directions are obtained with a $180^{\circ}$ rotation
of the outgoing wave directions.} instead of the outgoing one to better
visualize the model impact of waves\vadjust{\goodbreak} on the coast during storms. From
Figure~\ref{krigrd} we can see that the model captures local features
of the wave field, that arrows rotate in the upper part of the plot
(following the contour of the Gargano promontory and the presence of
the outliers), and that more variable estimates are obtained near the
coast line (mimicking similar behavior of wave heights before calibration).
Finally, we compare our results with those from the nonspatial model
(which we ran for 20,000 iterations, keeping the second half for
estimation, using priors as described in Section~\ref{modfitind}). The
estimated posterior mean direction is 0.5507 ($[0.3278, 0.7810]$ 95$\%$
C.I.) and the posterior concentration is 0.8272 ($[0.7748, 0.8667]$
95$\%$ C.I.). The leave-one-out validation yielded an average
predictive error of 0.1553, revealing that the spatial model yielded a
reduction of 68\%.

%s5 #&#
\section{Summary and future work}\label{sec5}

We have introduced a class of spatial process models which can be used
to study circular data that is anticipated to have spatial structure.
In particular, adopting the class of wrapped distributions for
describing circular data, we have shown how a usual Gaussian process
specification can be converted to a wrapped normal process
specification. We have demonstrated that fitting for such models can be
done straightforwardly within a hierarchical modeling framework using
MCMC methods with adaptive truncation. We have also shown how to
implement kriging for such models. Our motivating application has
revealed the predictive benefit of the spatial modeling. %how to
%introduce measurement error into these models and finally,

Particularly with directional data, one would expect there to be
concerns regarding measurement error, for instance, with monitors
recording wind direction and with buoys measuring wave direction. We
are unaware of any measurement error work in the context of directional
data but addressing it using wrapped Gaussian processes turns out to be
straightforward. It will be captured by a \textit{nugget} similar to the
usual geostatistical modeling setting. In fact, a frequent
interpretation of the nugget is measurement error [\citet
{banerjeebook}]. Suppose the observed angular data are collected with
conditionally independent measurement error, that is, $X_{o,j} \sim
\operatorname{WN}(Y_{t,j}, \tau^{2}),$ where $Y_{t,j} =X_{t,j}+ 2\pi K_{t,j}$ and
$X_{t,j}$ is the true angular direction, with independence across $j$.
As above, we introduce latent $K_{o,j}$ and model the joint
distribution of $X_{o,j}, K_{o,j}$ as $N(X_{o,j}+ 2\pi K_{o,j}|X_{t,j},
\tau^2)$. The latent true directions follow the wrapped GP of Section
\ref{WSGP}. That is,
\[
\mathbf{X}_{t}=\bigl(X_{t}(s_{1}),X_{t}(s_{2}),
\ldots,X_{t}(s_{p})\bigr) \sim \operatorname{WN}\bigl(\mu \mathbf{1},
\sigma^{2}\mathbf{R}(\phi)\bigr),
\]
where $R(\phi)_{ij}= \rho(s_{i} - s_{j};\phi)$. Introducing $\mathbf
{K}_{t}$, the joint distribution for $\mathbf{X}_{t}, \mathbf{K}_{t}$
takes the form $N(\mathbf{X}_{t}+ 2\pi\mathbf{K}_{t}|\mu\mathbf
{1},\sigma^{2}\mathbf{R}(\phi))$.
The full model takes the form
%
%e13 #&#
\begin{equation}
\Pi_{j} \bigl[X_{o,j}+2\pi K_{o,j}|X_{t,j},
\tau^{2}\bigr] \bigl[\mathbf{X}_{t} +2 \pi
\mathbf{K}_{t}|\mu, \sigma^{2}\mathbf{R}(\phi)\bigr] [\mu]
\bigl[\sigma^2\bigr] \bigl[\tau^2\bigr] [\phi].
\label{MEGP}
\end{equation}

An issue that requires further investigation is the matter of the
assumption of stationarity in the covariance function. It would be
useful to develop diagnostics to examine this, paralleling those for
linear spatial data. At present, all we can suggest is model comparison
using average predictive error as in Section~\ref{examples}.

Future work will investigate promise of extending the wrapped normal
process to a wrapped $t$-process through the usual Gamma mixing that
extends a GP to a $t$-process [see, e.g., \citet{zhang07}]. Extension to
the $t$-process discussed in \citet{heyden05} would be more challenging.
It will also lead us to incorporate dynamic structure into our
modeling; with regard to our data set, we have wave direction
information at various temporal resolutions.
Finally, we will explore two data assimilation issues. The first is to
fuse the angular data produced by the WAve Model (WAM) with the buoy
data (RON) to improve our interpolation of wave direction. The second
involves joint modeling of the wave direction data with the available
associated wave height data. Hopefully, joint modeling will enable a
version of co-kriging to improve the individual interpolations.

\section*{Acknowledgments}
The authors thank the Coastal Defense Unit, ISPRA, Italy, for providing
the data. They are grateful to Daniela Cocchi and Clarissa Ferrari for
many discussions and useful suggestions on this topic. The authors wish
to thank the reviewers and the editor for very useful comments and
suggestions that helped to improve considerably the paper.

%suskaldyti doi
%
% imsref loaded by akundreckaite, 2012-07-05 10:21:55

\printaddresses

\end{document}